\documentclass[graybox]{svmult}

\usepackage{mathptmx}       
\usepackage{helvet}         
\usepackage{courier}        
\usepackage{type1cm}        
\usepackage{makeidx}         
\usepackage{graphicx}        
\usepackage{multicol}        
\usepackage[bottom]{footmisc}

\makeindex             

\begin{document}
\newcommand{\mnras}{MNRAS}
\newcommand{\apj}{ApJ}
\newcommand{\apjl}{ApJL}
\newcommand{\aap}{A\&A}
\newcommand{\apjs}{ApJS}
\newcommand{\pasp}{PASP}
\newcommand{\nat}{Nature}

\title*{Gravity at Work: How the Build-Up of Environments Shape Galaxy Properties}
 \titlerunning{Gravitational Heating} 
\author{Sadegh Khochfar}
\institute{Sadegh Khochfar \at Max-Planck-Institute for Extraterrestrial Physics, Giessenbachstrasse 1, 85748 Garching, Germany \email{sadeghk@mpe.mpg.de}}
%
%
\maketitle

\vskip-1.2truein

\abstract{We present results on the heating of the inter-cluster medium (ICM) by gravitational potential energy from in-falling satellites. We calculate the available excess energy of baryons once they are stripped from their satellite and added to the ICM of the hosting environment. this excess energy is a strong function of environment and we find that it can exceed the contribution from AGNs or supernovae (SN) by up to two orders of magnitude in the densest environments/haloes. Cooling by radiative losses is in general fully compensated by gravitational heating in massive groups and clusters with hot gas temperature $> 1 $ keV.  The reason for the strong environment dependence is the continued infall of substructure onto dense environments during their formation in contrast to field-like environments. We show that gravitational heating is able to reduce the number of too luminous galaxies in models and to produce model luminosity functions in agreement with observations.    }

\section{Introduction}
\label{sec:1}
Within the $\Lambda$CDM paradigm of structure formation, the growth of galaxies is governed by dissipational processes such as cooling of gas and associated star formation, as well as dissipationless processes such as mergers of stellar systems that are already in place e.g. \cite{1991ApJ...379...52W}. The interplay of such processes regulates the properties of individual galaxies. At early times  during the cosmic evolution $(z \geq 2$), when cooling times are short in low mass  \cite{1993MNRAS.264..201K} and  massive haloes \cite{2009Natur.457..451D,2009ApJ...700L..21K} dissipational processes associated with star formation either is disc-like systems or gas rich mergers \cite{2010arXiv1007.1463K} dominate the growth of galaxies. With the onset of downsizing in the star formation rate of galaxies at low redshifts e.g. \cite{2007ApJ...660L..43N} this picture changes, in particular for massive early-type galaxies that can accrete up to $80 \%$ of their stellar mass by $z=0$ from satellites  \cite{2006MNRAS.370..902K}. The importance of such 'dry' mergers for the growth of massive galaxies at late times has been investigated e.g. in \cite{2003ApJ...597L.117K,2006ApJ...636L..81N,2009MNRAS.397..506K} and recently linked to the size-evolution of massive early-type galaxies  \cite{2006ApJ...650...18T,2006ApJ...648L..21K,2009ApJ...699L.178N}.  

Feedback is generally assumed to play an important role in trends as the ones mentioned above e.g. \cite{2010MNRAS.402.1536S}. In particular supernovae feedback has been considered in shaping the faint-end of the luminosity function \cite{2003ApJ...599...38B,2007ApJ...668L.115K}. At the luminous end the situation appears more complicated with several competing effects at play as e.g. AGN-feedback \cite{2006MNRAS.370..645B} or gravitational heating by in-falling sub-structure \cite{2008ApJ...680...54K,2009ApJ...697L..38J}. 

The observed properties of the galaxy population in high-density environments show distinct properties in contrast to the field galaxy population. The population of spiral galaxies is gas-poor \cite{2006ApJ...641L..97M} and has smaller HI discs \cite{2004cgpc.symp..305V} compared to spiral galaxies in the field. Besides affecting the properties of individual galaxies, the morphological mix of the population as whole changes in high density environments as well. The so-called density-morphology relation shows and increasing trend in the  fraction of early-type galaxies as a function local density  \cite{ad80}.  The process of ram-pressure stripping \cite{1972ApJ...176....1G} is able to remove gas from satellite galaxies orbiting in high density environments, and truncate disc sizes.  The change in morphological fraction is generally attribute to a higher fraction of mergers during the formation epoch of the high density environment.

While the above mentioned processes {\it actively} change individual galaxies we here address how the build-up of the environment {\it passively} affects the state of the central galaxy population living in it via the ability/inability to cool gas. 
Numerical simulations show that shock heating of gas falling into the cluster potential is able to heat gas initially to the virial temperature of the cluster \cite{1999ApJ...525..554F}. In addition conversion of gravitational potential energy of orbiting satellites is able to heat the inter-cluster medium (ICM) \cite{2008ApJ...680...54K,2008MNRAS.383..119D}. In the following we will focus on the latter effect and include it into a semi-analytic model (SAM) to estimate its impact on the galaxy population.  

\section{Model}
The results we present here are based on the SAM described in detail in \cite{2005MNRAS.359.1379K,2008ApJ...680...54K}, using the three-year WMAP cosmology with $\Omega_0=0.27$, $ \Omega_\Lambda=0.73$ ,$ \Omega_b/\Omega_0=0.17$, $\sigma_8=0.77$, $h=0/71$ \cite{2007ApJS..170..377S}. We briefly summarize here the main physical models that are important in high density environments, and that we have included in our SAM.

\subsection{Shock Heating}
Shock-heating of satellite gas takes place when a small satellite falls within the large potential well of a host dark matter halo and starts interacting with the hot halo gas. Generally SAMs assume that the shock heating is taking place instantaneously, thereby removing all hot halo gas from satellites and adding it to the hot gas reservoir of the central galaxy \cite{1993MNRAS.264..201K}. It has been that this process, if modeled instantaneously might be too efficient resulting in too many passive satellite galaxies in high density environments. \cite{2009MNRAS.394.1213W}. We adopt a simple prescription  for the process  of shock heating  assuming that the rate at which mass is shock heated from the satellite is related to the halo dynamical time $t_{dyn}$ via $\dot{M} \sim M/t_{dyn}$. This is effects allows the satellite to hold on to his hot gas for a longer period of time and to continue cooling.

\subsection{Ram Pressure Stripping} 
Following \cite{1972ApJ...176....1G}, we assume that gas is stripped from the satellite once the dynamical pressure is able to over- come the gravitational force binding the gas to the satellite. In terms of energy deposited within the satellite gas, this can be approximated by
\begin{equation}
\dot{E}_{ram}=\rho_{hot} v^3_{\bot}\pi r_h^2
\end{equation}
Here, we take $v_{bot}$ as the velocity of the satellite perpendicular to its disc orientation and assume that the orbital velocity of the satellite is comparable to the sound speed $c_{gas}$ of the hot gas. The efficiency of this process for a face-on disk should be maximal, and the efficiency for an edge-on disk should be minimal; we take this into account by assuming that the disk orientation is random with respect to the infall direction. This is in good agreement with cosmological dark matter simulations that show that the spin vectors of merging dark matter halos are randomly aligned to each other \cite{2006A&A...445..403K}. For simplicity, we calculate the density $\rho_{hot}$ by taking the average density of hot gas within the host haloÕs virial radius, and take $r_h$ to be the characteristic half-mass radius of the gas within the satellite. The rate of stripped material is then calculated dividing $ \dot{E}_{ram}$ by the specific energy of the gas on which it is acting.

\subsection{Gravitational Heating}
The energy needed to strip baryonic material from satellites is provided by the conversion of gravitational potential energy, that the satellite gains during its infall on  a slightly bound $E_{bind} \sim 0$.  Parabolic orbits have indeed been shown to be the most probable based on cosmological  N-body simulations \cite{2006A&A...445..403K}. Gas that it stripped from a satellites can in principle have potential energy left, after it has been used up for the stripping process, and we assume here that this energy will be used to heat the inter-cluster medium. In practice we apply following equation to calculate the amount of gravitational potential energy left to heat the ICM:
\begin{equation}\label{grav}
\dot{E}_{grav}=\sum_{i=1,n_{sat}} \dot{M}_{gas,i}\left(\Delta \phi - \Delta E_{strip} \right).
\end{equation}
The amount of stripped gas for each satellite $i$ orbiting within the host halo is $\dot{M}_{gas,i}$ and the total energy needed to strip the material from the satellites is  $\Delta E_{strip}$. The heating of the ICM based on Eq. \ref{grav} predicts that for the same sum of stripped material from massive satellites, heating is less efficient than if that material would be stripped from less massive satellites.  
 \section{ Results}
\begin{figure}[tb]
\includegraphics[scale=.50]{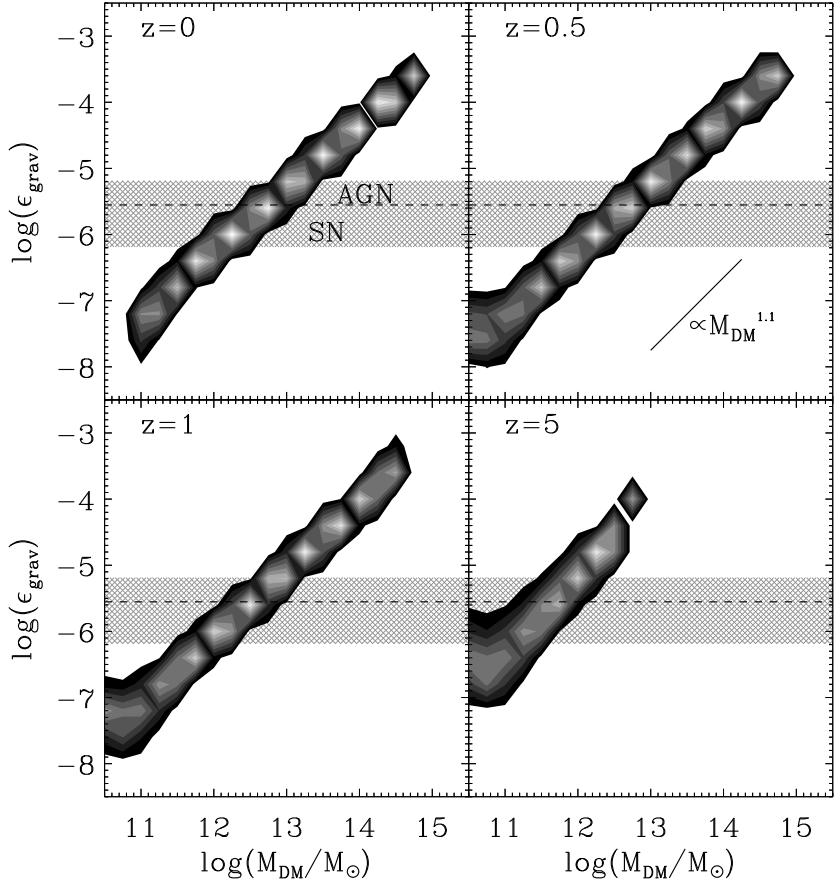} 
\caption{The integrated efficiency of gravitational heating $\epsilon_{grav}$ as a function of host halo mass. Galaxies living the the most massive haloes always experience  the largest contribution from gravitational heating, due to the continued infall from satellites. Shaded areas show the expected range in efficiencies for SN and AGN-feedback. }
\label{fig1}
\end{figure}
The time integral of Eq. \ref{grav} over the evolution of a high density environment can be quite substantial and, if	expressed	 in terms of $E_{grav,tot}=\epsilon_{grav} m_* c^2$, we find values for $\epsilon_{grav}$ ranging from a few time $10^{-8}$ to a few times $10^{-4}$ in galaxies with stellar masses $M_* \sim 10^{10}$ and $\sim 5 \times 10^{12}$ M$_{\odot}$, respectively. To emphasize that this trend is driven by the environment, we show in Fig. \ref{fig1} the trend with halo mass which we take as a proxy for the environment. Massive haloes have more gravitational heating during their evolution than less massive ones.  One can understand this behavior by considering the accretion of satellites onto halos. In general, the most massive halos at any redshift had the largest accretion rates in the past, which explains the large amount of gravitational heating and its dependence on halo mass. It is worth comparing gravitational heating to other common heating mechanisms such as supernovae and AGNs. Comparing $\epsilon_{grav}$ to $\epsilon_{SN} \sim 2.8 \times 10^{-6}$ and $\epsilon{BH} \sim 6.5 \times 10^{-6} - 6.5 \times 10^{-7}$ shows that in general, gravitational heating is more efficient than supernova feedback only in galaxies larger than a few times $10^{11}$ M$_{\odot}$ and in halos more massive than $5 \times 10^{12}$ M$_{odot}$ at  z = 0. This regime corresponds to massive field galaxies and extends into group-like environments. For even more massive galaxies and dark halo masses larger than $10^{13}$ M$_{\odot}$, gravitational heating starts dominating over proposed AGN feedback rates. 
 \begin{figure}[tb]
\includegraphics[scale=.50]{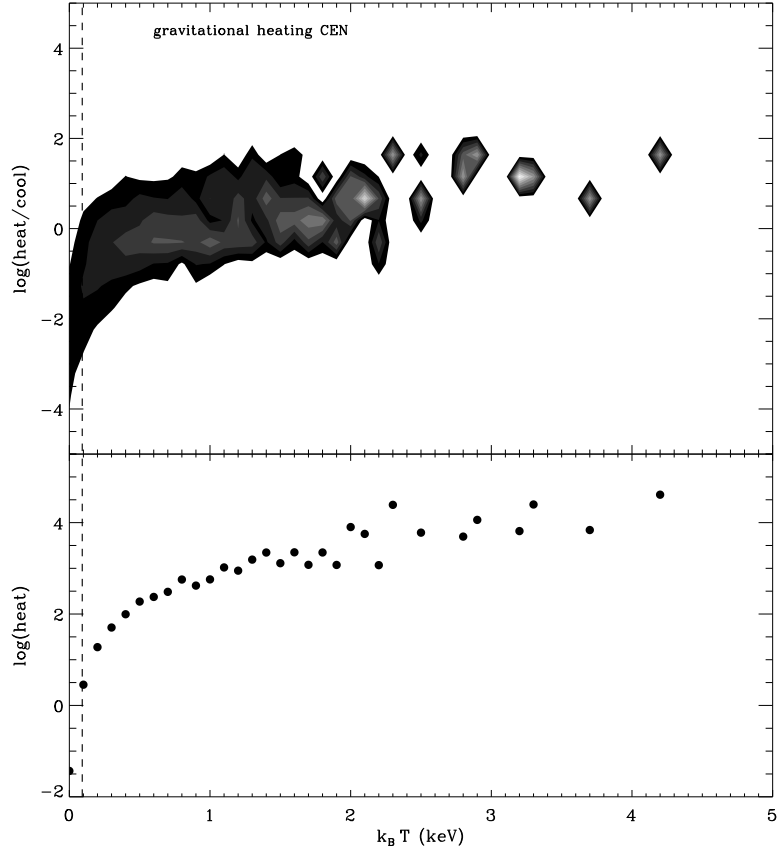}
\caption{ Top: the probability distribution for the ratio of heating to cooling rates as function of environment. The x-axis shows the temperature of the hot halo gas as a proxy for environment. Bottom: The average gravitational heating rate as a function of  environment.}
\label{fig2}
\end{figure}

To illustrate the contribution from gravitational heating, we display the contours of the conditional probability for the ratio of heating to cooling that individual galaxies experience in a given environment. The top panel in Fig. \ref{fig2} shows the probability contours for central galaxies at $z = 0.1$ in our simulations. We translate the deposited energy per unit time into a heating rate, labeled {\it heat} in Fig. ref{fig2}, by calculating the amount of cold gas that can be heated to the virial temperature of the dark matter halo in which the galaxy resides. The cooling rate, labelled {\it cool} is the standard radiative cooling rate that we calculate in our SAM based on the prescription in \cite{2005MNRAS.359.1379K}.
Figure \ref{fig2} shows clearly that gravitational heating dominates in environments with temperatures above $\sim $keV. The heating rate for the central galaxies can be up to $10^2$ times larger than the cooling rate, and in the most dense environments the heating rate becomes $10^4$ M$_{\odot}$ yr$^{-1}$. The heating rate shows a clear environmental dependence that reflects the higher abundance of satellites that contribute to gravitational heating. From these results, one expects that star formation in central galaxies of dense environments will be shut down. 

\begin{figure}[tb]
\includegraphics[scale=.50]{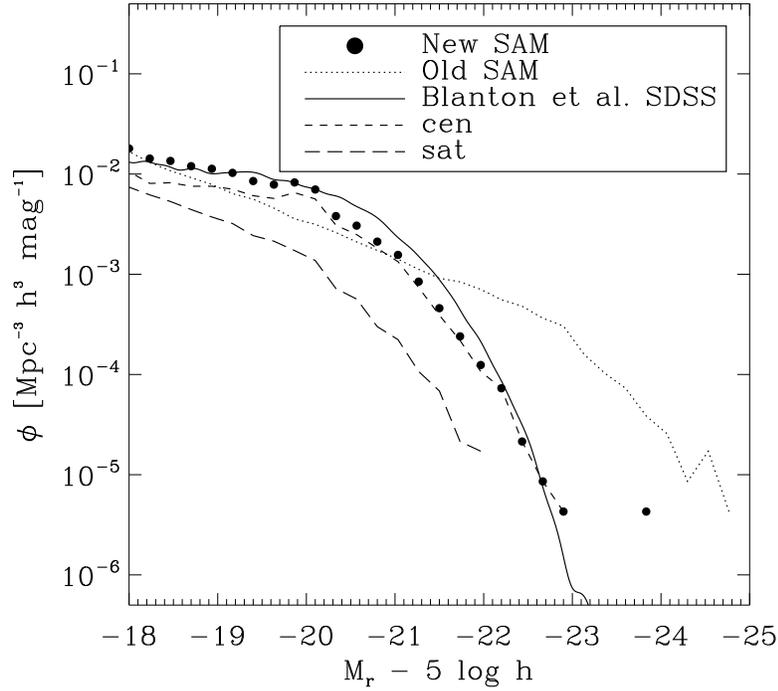} 
\caption{ The $r-$band luminosity function of galaxies in our model including gravitational heating (filled circles) and without (dotted line), compared to the SDSS results (solid line) by \cite{2003ApJ...592..819B}  }
\label{fig3}
\end{figure}

To further illustrate the effect of gravitational heating on the galaxy population we show in Fig. \ref{fig3} the luminosity function of galaxies at $z\sim 0$. We find a good match to the observed one by \cite{2003ApJ...592..819B}. The effect of gravitational heating is strongest for massive galaxies. These tend to live predominantly in high density environments, subject to gravitational heating. For comparison we also show a model without the contribution of gravitational heating.     
This model severely overproduces the number density of luminous galaxies. 

\section{Conclusions}
We here show the effects that the surplus of gravitational potential energy from in-falling baryonic matter can have on the heating of the ICM. Galaxies in low density environments are generally not affected by gravitational heating as the contribution is small compared to other feedback sources like SN or AGN feedback. We find that in the most dense environments, like massive groups or clusters, the constant infall of substructure is able to provide enough energy to level, and even exceed the losses due to radiative cooling, hence stopping any cooling in such environments. The integrated contribution from gravitational heating is in cluster environments clearly larger than the contribution from AGNs suggesting that this could be an important feedback source. This conclusion is supported by looking at the luminosity function of galaxies. Including gravitational heating is able to reduce the amount of too luminous galaxies and to bring them in agreement with observations. To further test the importance of gravitational heating high-resolution simulations of galaxy clusters are required. Such simulations should be able to reveal self-consistently the importance of the effects discussed here.

\begin{acknowledgement}
SK would like to thank the Anna Pasquali and Ignacio Ferreras for organizing a stimulating meeting at the JENAM 2010.
\end{acknowledgement}
%


\end{document}